\documentclass[english,aps,floats,twocolumn,showpacs,nofootinbib]{revtex4}
\usepackage{pslatex}
\usepackage[T1]{fontenc}
\usepackage[latin1]{inputenc}
\usepackage{graphicx}
\usepackage{epsfig}
{\newcommand{\lsim}{\mbox{\raisebox{-.6ex}{~$\stackrel{<}{\sim}$~}}}
{\newcommand{\gsim}{\mbox{\raisebox{-.6ex}{~$\stackrel{>}{\sim}$~}}}
\newcommand{\be}{\begin{equation}}
\newcommand{\ee}{\end{equation}}
\newcommand{\bea}{\begin{eqnarray}}
\newcommand{\eea}{\end{eqnarray}}

\begin{document}
\title{Extended Zee model for Neutrino Mass, Leptogenesis and Sterile Neutrino like Dark Matter}
\author{Narendra Sahu}
\email{n.sahu@lancaster.ac.uk}
\affiliation{Cosmology and Astroparticle Physics Group, University of
Lancaster, Lancaster LA1 4YB, UK}
\author{Utpal Sarkar}
\email{utpal@prl.res.in}
\affiliation{Theory Division, Physical Research Laboratory,
Navarangpura, Ahmedabad, 380 009, India}
\begin{abstract}

We propose an extension of the standard model with a $U(1)_{\rm B-L}$
global symmetry that accommodates radiative neutrino masses along with
dark matter and leptogenesis. The observed matter antimatter asymmetry of 
the universe is generated through the leptogenesis route keeping the 
$U(1)_{\rm B-L}$ symmetry intact. The $B-L$ global symmetry is then 
softly broken, providing the sub-eV neutrino masses. The model then 
incorporates a MeV scale sterile neutrino like dark matter.

\end{abstract}
\pacs{12.60.Fr, 14.60.St, 95.35.+d, 98.80.Cq}
\maketitle

\section{Introduction}
In recent times, there have been growing interests on models of neutrino mass, dark
matter and baryon asymmetry~\cite{sahu&sarkar,neutrino_dark_lep}. While the low energy neutrino
oscillation data~\cite{solar_data,atmos_data,kamland_data} indicate that at least
two of the physical left-handed (LH) neutrinos have tiny masses and therefore they mix
among themselves, the galactic rotation curve~\cite{rotation_curve}, gravitational
lensing~\cite{lensing} and large scale structure~\cite{structure} strongly demand
that there exist non-baryonic dark matter in the present universe, which has only
matter (visible plus dark) components. The antimatter components in the present
universe are vanishingly small. This implies that the asymmetry between matter and
antimatter components is maximal today. Currently this asymmetry has been precisely
measured by the Wilkinson Microwave Anisotropy Probe (WMAP)~\cite{wmap}
and is given by
\begin{equation}
\left( \frac{n_B - n_{\overline{B}}}{n_\gamma} \right)_0 \equiv \left( \frac{n_B}{n_\gamma}
\right)_0 = 6.1^{+0.3}_{-0.2} \times 10^{-10}\,,
\end{equation}
where $n_B$ is the baryon density and $n_\gamma$ is the primordial photon density.

Despite the success of the standard model (SM), it can not explain any of the above
phenomena: non-zero neutrino mass, existence of dark matter and matter--antimatter asymmetry 
of the universe. Explanation of all of these phenomena requires an extension of SM. The
observed neutrino oscillation experiments deal only with
the mass square differences of the neutrinos, and hence, they leave an ambiguity on
the nature of neutrinos to be either Dirac or Majorana. If the neutrinos are assumed
to be Majorana particles, the sub-eV neutrino masses can naturally be generated through the
celebrated seesaw mechanisms: type-I~\cite{typeI_group} and type-II~\cite{typeII_group}.
In either case, the neutrino mass is suppressed by the scale of lepton (L) number
violation. The smallness of the neutrino mass could also be
explained naturally without invoking any large lepton number violating scale by the radiative
mechanisms~\cite{radiative_models}. The loop factor then introduces the suppression
required to keep the neutrino masses small. However, these models have the generic
problem of explaining the baryon asymmetry of the universe, since the fast lepton number
violation required to generate the neutrino masses, also erases any baryon asymmetry of
the universe~\cite{maraidalsarkar.99}. In this note we aim to solve this problem and
generate the neutrino masses and leptogenesis simultaneously in a radiative model. The
lepton asymmetry is generated without any B-L violation, similar to leptogenesis in models 
of Dirac neutrinos \cite{lindner,ghs7}. The present model also accommodates a sterile neutrino 
like dark matter candidate. In contrast to the generic radiative neutrino mass models, where 
the masses of the fields propagating in the loop are expected
to be at the TeV scale, in the present model the masses of the fields propagating in the
loop could be as high as $10^{10}$ GeV or so. As a result a leptogenesis could be possible
from the out-of-equilibrium decay of these heavy charged scalars in the early universe.

The paper is arranged as follows. In section II we propose a radiative neutrino mass
model which has an exact B-L symmetry to begin with. The B-L symmetry is then softly
broken to give rise neutrino masses. Section III is devoted to estimate neutrino masses
originating from the soft B-L violation. In section IV we calculate lepton asymmetry
from a conserved B-L symmetry. In section V we estimate the lifetime of a decaying 
sterile neutrino like dark matter. Finally section VI concludes.

\section{Extended Zee Model with conserved B-L symmetry}

Within the SM neutrinos are massless. This is because $B-L$ is an exact symmetry
of SM. So any radiative mechanism within SM can not give rise to neutrino masses.
One of the simplest extensions of SM is Zee model where SM is extended with a charged
scalar, say $\eta^-$, and a second doublet Higgs, say $\phi_b$, to generate neutrino
masses through one loop radiative correction. The antisymmetric couplings of $\eta^-$
to Higgses ($\mu_{ab} \eta^- \phi_a \phi_b$) and leptons ($f_{ij}\eta^- \ell_i \ell_j$)
then together violate lepton number by two units and neutrinos acquire
masses through one loop radiative correction. Since the couplings of the
Higgses and the charged leptons to the charged scalar is antisymmetric,
the diagonal elements of the neutrino mass matrix vanishes identically
in a flavor basis. The lepton number violating couplings that appear
in the neutrino masses, also give rise to lepton number violating
interactions in the early universe. The observed neutrino masses would
then imply that the lepton number violating interactions in the early
universe is too fast, which will wash out~\cite{maraidalsarkar.99} any matter--antimatter
asymmetry of the universe in the presence of the sphalerons before the
electroweak phase transition.

We propose an extension of the Zee model, which can simultaneously explain
the observed neutrino masses, the baryon asymmetry of the universe and
also provide a candidate for dark matter of the universe. We
extend the Zee model with an additional charged scalar $\chi^-$ and
three sterile neutrino like fields $N_L$, and impose a $U(1)_{B-L}$ global symmetry.
The particle content of the model and their quantum numbers are presented in 
table \ref{table-1}.
\begin{table}[h]
\begin{center}
\caption{Particle content and their quantum numbers.\label{table-1}}
\begin{tabular}{|c|c|c|}\hline
Particle Content & $SU(3)_C\times SU(2)_L\times U(1)_Y$ &  $U(1)_{B-L}$\\ \hline
$\ell_L$  & (1,2,-1) & -1 \\[2mm] \hline
$e_R^-$ & (1,1,-2) & -1 \\[2mm] \hline
$\phi_a$, $\phi_b$  & (1,2,1) & 0\\[2mm] \hline
$\chi^-$   & (1,1,-2) & -2 \\[2mm] \hline
$\eta_a^-$ & (1,1,-2) & 0\\[2mm] \hline
$N_L$ & (1,1,0) & -1\\[2mm] \hline
\end{tabular}
\end{center}
\end{table}
The relevant terms in the Lagrangian are then given as:
\begin{eqnarray}
{\cal L} & \supseteq & M_\chi^2 \chi^\dagger \chi + M_\eta^2 \eta^\dagger \eta + f_{ij} \chi^\dagger 
\ell_{iL} \ell_{jL} + (\mu)_{a b} \eta \phi_a \phi_b \nonumber\\
&& + h_{ij} \eta^\dagger \overline{N_{iL}} e_{jR} + Y^a_{ii} \overline{\ell_{iL}} \phi_a e_{iR}
+h.c.\, 
\label{Lagrangian}
\end{eqnarray}
where $i,j=e, \mu, \tau$ are family indices and $a,b$ corresponds to two Higgs doublets. Since
$N_L$ is electrically neutral it can have Majorana masses $M_{N_L}$. In Eq. (\ref{Lagrangian}) 
we have assumed that $\phi_a$ couples only to leptons and $\phi_b$ couples only to 
quarks~\cite{petcov} apart from their self-interactions. However, this can be achieved by 
using an additional $Z_2$ symmetry. We also assume that the couplings of leptons to $\phi_a$ 
is diagonal similar to ref.~\cite{petcov}. As a result there is no tree level flavor changing 
processes induced by $Y_{ij}$ in the lepton sector.

As the fields $\chi^-$ and $\eta^-$ are charged they can not acquire any vacuum expectation value
(VEV). The scalar potential involving the $\phi_a$ and $\phi_b$ then can be given as:
\begin{eqnarray}
V(\phi_a, \phi_b) &=& -M_a^2 |\phi_a|^2 -M_b^2 |\phi_b|^2 + M_{ab}^2 \phi_a^\dagger \phi_b+ 
\lambda_a |\phi_a|^4 \nonumber\\
&& + \lambda_b |\phi_b|^4 + \lambda_{ab}|\phi_a|^2 |\phi_b|^2 + \lambda'_{ab} 
|\phi_a^\dagger \phi_b|^2 \nonumber\\
&& + \frac{\lambda''_{ab}}{2} [(\phi_a^\dagger \phi_b)^2+h.c.]
\end{eqnarray}
where $M_a^2, M_b^2 >0$. The stability of the potential requires $\lambda_a, \lambda_b > 0$
and $\lambda_{a b} > -\sqrt{\lambda_a \lambda_b}$. The VEV of Higgses $\phi_a$ and
$\phi_b$ can be given as
\begin{equation}
\langle \phi_a \rangle = v_a \;\;\; {\rm and} \;\;\; \langle \phi_b \rangle =v_b
\end{equation}
Then the two VEVs are related by
\begin{equation}
v=\sqrt{v_a^2+v_b^2}=174 {\rm GeV}\,,\;\;\; v_a=v \sin \beta \;\;\; {\rm and} \;\;\; v_b=v \cos \beta\,,
\end{equation}
where $\beta=\tan^{-1}(v_a/v_b)$. 

\section{Soft B-L Violation and Radiative Neutrino Masses}

The $U(1)_{\rm B-L}$ symmetry is allowed to be broken softly by
\begin{equation}
\mathcal{L}_{\rm soft}=m^2 \eta^\dagger \chi + h.c.\,.
\end{equation}
As a result there is a mixing between $\eta^-$ and $\chi^-$. In the flavor basis the mass 
matrix of $\eta^-$ and $\chi^-$ is then given by
\begin{equation}
\mathcal{M}^2=\pmatrix{ M_\eta^2 & m^2 \cr
m^2 & M_\chi^2 }\,.
\end{equation}
Diagonalising the above mass matrix we get the eigenvalues: 
\begin{equation}
M^2_{\eta',\chi'}=\frac{ (M_\eta^2+M_\chi^2) \pm \sqrt{(M_\eta^2-M_\chi^2)^2+4 m^4}}{2}
\end{equation}
corresponding to the mass eigenstates $\eta'=\cos \theta \eta^- + \sin \theta \chi^-$ and 
$\chi'=\cos \theta \chi^- - \sin \theta \eta^-$, where 
\begin{equation}
\theta=\frac{1}{2}\tan^{-1} \left( \frac{2 m^2}{M_\chi^2-M_\eta^2}\right)\,.
\end{equation}

\begin{figure}[htbp]
\begin{center}
\epsfig{file=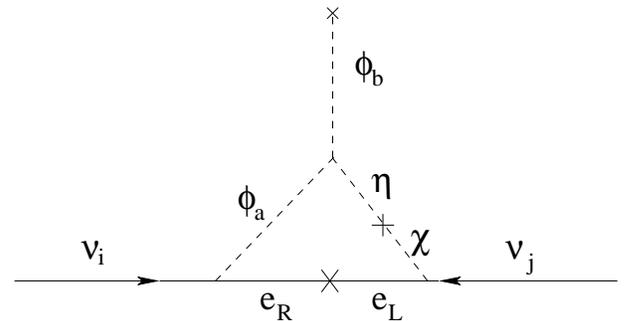, width=0.45\textwidth}
\caption{One loop radiative correction diagram for neutrino masses arises through
the mixing between $\eta$ and $\chi$.\label{neutrino_mass}}
\end{center}
\end{figure}

Through the mixing between $\eta^-$ and $\chi^-$ lepton number is violated
by two units. As a result neutrino mass is generated through the one loop radiative
correction diagram as shown in fig. \ref{neutrino_mass}. 
From fig. \ref{neutrino_mass}, the neutrino mass can be estimated to be
\begin{equation}
\left( M_\nu \right)_{ij} = \left( Y_{ii}^a M_i- Y_{jj}^a M_j \right) f_{ij} \left( m^2 \mu_{ab} v_b \right). |I|
\label{neutrinomass}
\end{equation}
where $M_i$ and $M_j$ are the diagonal charged lepton mass matrix and the integral: 
\begin{eqnarray}
I &=& \int \frac{d^4 q}{(2\pi)^4} \frac{1}{q^2-M_\chi^2} \frac{1}{q^2-M_\eta^2}
\frac{1}{q^2-M_a^2} \frac{1}{q^2}\nonumber\\
&=& \frac{i}{16 \pi^2} \left[ \frac{ M_a^2\ln (M_\chi^2/M_\eta^2) + M_\chi^2 \ln (M_\eta^2/M_a^2)
+ M_\eta^2 \ln (M_a^2/M_\chi^2)}{(M_\chi^2-M_\eta^2)(M_\chi^2-M_a^2)(M_\eta^2-M_a^2)}\right] \nonumber\\
\end{eqnarray}
From Eq. (\ref{neutrinomass}) it is clear that $M_\nu$ is symmetric with respect to the family 
indices $i$ and $j$ and the diagonal elements of $M_\nu$ are also zero. This is because of 
our assumption about the coupling of $\phi_a$ to the leptons. However, in general, this is 
not true.  

If we assume that $M_\eta, M_\chi \gg M_a, M_b$, then the neutrino mass is simply
\begin{eqnarray}
\left( M_\nu \right)_{ij} &\simeq &  \frac{1}{16 \pi^2} \left( M_i^2 - M_j^2 \right) f_{ij}\frac{\mu_{ab}  
m^2 \cot \beta}
{(M_\chi^2 - M_\eta^2)}\nonumber\\
&& \times \left[\frac{\ln (M_\eta^2/M_b^2)}{M_\eta^2}- \frac{ \ln (M_\chi^2/M_b^2)}
{M_\chi^2} \right] \nonumber\\ 
&=& M_0 F_{ij}\,,
\end{eqnarray}
where 
\begin{equation}
M_0 = \frac{1}{16 \pi^2} \frac{M_\tau^2 \mu_{ab} m^2} 
{(M_\chi^2 - M_\eta^2)}\cot \beta \left[\frac{\ln (M_\eta^2/M_a^2)}{M_\eta^2}- \frac{ \ln (M_\chi^2/M_a^2)}
{M_\chi^2} \right]
\end{equation}
and $F_{ij}=f_{ij} \delta_{ij}$, where 
\begin{equation}
\delta_{ij}=\frac{M_i^2-M_j^2}{M_\tau^2}\,.
\end{equation}
We now estimate the magnitude of front factor $M_0$ by using the following sample set:  
\begin{equation}
M_\eta=3\times 10^{10} {\rm GeV}\,,\;\; M_{\chi}=5\times 10^{10} {\rm GeV}\;\;{\rm and}\;\; M_a=500 {\rm GeV}
\end{equation}
We also choose the mas dimension coupling to be $\mu=10^{15}$ GeV and the soft B-L violation scale 
$m=10^9$ GeV. Then we get the symmetric neutrino mass matrix
\begin{equation}
(M_\nu)_{ij} = 0.3 \; {\rm eV} \; \cot \beta \; f_{ij} \delta_{ij}\,. 
\end{equation}
The coupling constants $F_{ij}=f_{ij} \delta_{ij}$ can be appropriately chosen to explain the current 
neutrino oscillation data. Similar to the other neutrino mass models the coupling constants $F_{ij}$ 
are constrained by the flavor changing processes like $\mu \rightarrow e +\gamma$, 
$\tau\rightarrow \mu +\gamma$, {\it etc...}. 

Note that in contrast to the original Zee model here the charged scalars flowing through the 
loop are super heavy. Therefore, they remain in out-of-thermal equilibrium for a while in the 
early universe, and hence, can generate a net lepton asymmetry consistently as discussed 
below.

\section{Leptogenesis from a conserved $B-L$ symmetry}

As the universe expands the temperature of the thermal bath falls. As a result
$\eta^-$ will go out-of-thermal equilibrium below its mass scale. Note that
$\eta^-$ has gauge interaction: $\eta^- \eta^+ \rightarrow B_\mu B^\mu$, where
$B_\mu$ is the $U(1)_Y$ gauge field. However, for $M_\eta \gsim 10^{10}$ GeV the
gauge interactions will remain in out-of-thermal equilibrium for a decoupled
temperature, say, $T_D \simeq M_\eta/10$. The partial decay width of $\eta^- \rightarrow
\overline{N_{iL}}e_{jR}^-$ can be given as
\begin{equation}
\Gamma_\eta=\frac{1}{8\pi}|h_{ij}|^2 M_\eta\,,
\label{decay-rate}
\label{eta-decay}
\end{equation}
where the family index $i,j=e, \mu, \tau$. At a cosmic temperature $T\simeq M_\eta$, if 
$\Gamma_\eta$ fails to compete with the Hubble expansion parameter
\begin{equation}
H=1.67 g_*^{1/2} \frac{T^2}{M_{\rm pl}}\,,
\label{hubble}
\end{equation}
where $g_*=106$ is the number of relativistic degrees of freedom, then $\eta$ goes out-of-thermal 
equilibrium. From equations (\ref{decay-rate}) and (\ref{hubble}) then we find 
that the out-of-equilibrium decay of $\eta^-$ occurs for
\begin{equation}
M_\eta \gsim 2.77 \times 10^{10} {\rm GeV} \left( \frac{h_{ij}}{10^{-3}} \right)^2\,.
\end{equation}
\begin{figure}[htbp]
\begin{center}
\epsfig{file=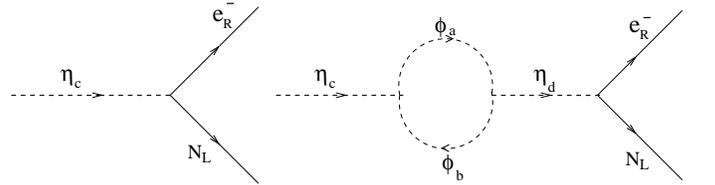, width=0.5\textwidth}
\caption{Tree level and one loop self energy correction diagrams whose
interference generates a net CP asymmetry.\label{leptogenesis_fig}}
\end{center}
\end{figure}
Note that decay of $\eta^-$ produce a pair of lepton and anti-lepton. Therefore, the 
decay of $\eta$ does not produce any lepton asymmetry. However, if there are 
at least two $\eta^-$ fields then there can be CP violation in the decay if $\eta$ 
fields. In the presence of their interactions, the diagonal mass $M_\eta^2$ in 
equation (\ref{Lagrangian}) is replaced by 
\begin{equation}
\eta_c^\dagger ({\mathcal M}_+^2)_{cd}\eta_d + (\eta_c^*)^\dagger ({\mathcal M}_-^2)_{cd} \eta_d^*
\end{equation}
where 
\begin{equation}
{\mathcal M}_{\pm}^2=\pmatrix{ M_{\eta_1}^2-i {\mathcal G}_{11} & -i{\mathcal G}_{12}^{\pm} \cr 
-i {\mathcal G}_{21}^{\pm} & M_{\eta_2}^2 - i {\mathcal G}_{22} }\,,
\label{eta-mass-matrix}
\end{equation}
where ${\mathcal G}_{cd}^+=\Gamma_{\eta_{cd}}M_{\eta_d}$, ${\mathcal G}_{cd}^-=\Gamma_{\eta_{cd}}^* M_{\eta_d}$, 
and ${\mathcal G}_{cc}=\Gamma_{\eta{cc}}M_{\eta_c}$ with $\Gamma_{\eta{cc}}\equiv \Gamma_{\eta_c}$. 
Similarly, the interaction term $\mu_{ab}\eta \phi_a \phi_b$ in equation (\ref{Lagrangian}) should be 
replaced by $(\mu_{ab})_c\eta_c \phi_a \phi_b \equiv \mu_c \eta_c \phi_a \phi_b$ and $(\mu_{ab})_d\eta_d 
\phi_a \phi_b \equiv \mu_d \eta_d \phi_a \phi_b$, where we have suppressed the symbols ``ab" in $\mu$. 
Now the absorptive part of one loop self energy diagram for $\eta_c \rightarrow \eta_d$ can be given by
\begin{equation}
\Gamma_{\eta_{cd}}M_{\eta_d}=\frac{1}{8\pi} \left( \mu_c \mu_d^* + M_{\eta_c} M_{\eta_d} \sum_{i,j}
h_{cij} h_{dij}^* \right)\,.
\end{equation}
Diagonalising  the mass matrix (\ref{eta-mass-matrix}) one will get two mass eigenvalues 
corresponding to the two mass eigenstates $\psi_1^\pm$ and $\psi_2^\pm$. Note that the mass 
eigenstates $\psi_1^+$ and $\psi_1^-$ (similarly $\psi_2^+$ and $\psi_2^-$) are not CP conjugate 
of each other even though they are degenerate mass eigenstates, while $\eta_1^+$ and $\eta_1^-$ 
(similarly $\eta_2^+$ and $\eta_2^-$) are CP conjugates of each other. Therefore, the decay of 
the lightest $\psi^\pm$, say $\psi_1^\pm$ can generate a net CP asymmetry through the interference 
of tree level and self energy correction diagram as shown in fig. (\ref{leptogenesis_fig}). The CP 
asymmetry is then given by~\cite{ma&sarkar}, 
\begin{equation}
\epsilon_1=\frac{ {\mathrm Im}\left[ (\mu_1 \mu_2^*) \sum_{ij} h_{1ij} 
h_{2ij}^* \right]}{16 \pi^2 (M_{\eta_2}^2-M_{\eta_1}^2)} \left[ \frac{M_{\eta_1}}
{\Gamma_{\eta_1}}\right]\,.
\label{cpasymmetry}
\end{equation}
Note that there is no radiative correction diagram. The decay of $\psi_1^\pm$ does 
not violate L-number, since the decay of $\eta^\pm$ does not violate lepton number. Therefore, 
the out-of-equilibrium-decay of $\psi_1^\pm$ does not produce any L-asymmetry. However, the 
decay of $\psi_1^\pm$, below its mass scale, generates an equal B-L asymmetry between $N_L$ 
and $e_R$ due to the CP violation~\cite{sahu&sarkar}. The B-L asymmetry stored in $e_R$ is 
then transferred to $e_L$ through the $t$-channel process $e_R e_R^c \leftrightarrow \phi_a^0 
\leftrightarrow e_L e_L^c$ as shown in the figure (\ref{fig2}).
\begin{figure}[htbp]
\begin{center}
\epsfig{file=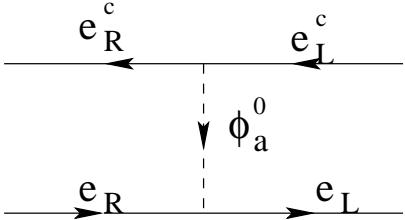, width=0.3\textwidth}
\caption{The L-number conserving process which transfer the B-L
asymmetry from right handed sector to the left-handed sector.}
\label{fig2}
\end{center}
\end{figure}
These interactions will remain in thermal equilibrium for all three
generations of charged leptons below $10^5$~GeV and hence there will
be an equal amount of $e_R$ and $e_L$ asymmetry. The B-L asymmetry
in $e_L$ will be converted to the baryon asymmetry of the universe
before the electroweak phase transition when the sphaleron processes are
in thermal equilibrium, while an equal asymmetry will remain in $N_L$. The 
two asymmetries will equilibrate with each other after the electroweak phase 
transition when the sphaleron processes go out-of-thermal equilibrium.   

The final baryon asymmetry thus generated can be given as
\begin{equation}
\frac{n_B}{s}\simeq \left( \frac{28}{79} \right) \frac{\epsilon_1}{g_* K (\ln K)^{0.6}}\,,
\end{equation}
where $K\equiv \Gamma_1/H$, $\Gamma_1$ being the decay rate of $\psi_1$, measures the 
deviation from equilibrium. If $K\ll 1$ then the baryon asymmetry is simply $(n_B/s)\sim 
\epsilon_1/g_*$. On the other hand, if $K>1$, then the final baryon asymmetry suffers from 
a suppression of $1/K (\ln K)^{0.6}$. If we assume that the CP violation is maximal, then 
by substituting $M_{\eta_1}^2/(M_{\eta_2}^2-M_{\eta_1}^2)\simeq {\cal O}(1)$ and $\mu_2/\mu_1
\simeq {\cal O} (1)$ in equation (\ref{cpasymmetry}) we get a CP asymmetry $\epsilon_1\sim 
10^{-6}$ for $h_{1ij}\simeq h_{2ij}\sim 10^{-3}$. Smaller values of $\epsilon_1$ can be 
conspired if we assume non-maximal CP violation. If we further assume that $K=1.0\times 10^3$, 
then the suppression factor will be $3\times 10^{-4}$. As a result we will get a net baryon 
asymmetry $n_B/s \sim 10^{-10}$. The observed baryon asymmetry is then given by $\eta=
n_B/n_\gamma=7 (n_B/s)$. An exact value of the suppression factor can be obtained by 
solving the required Boltzmann equation numerically which is beyond the scope of this paper.

\subsection{Washout Constraints}
\begin{figure}[htbp]
\begin{center}
\epsfig{file=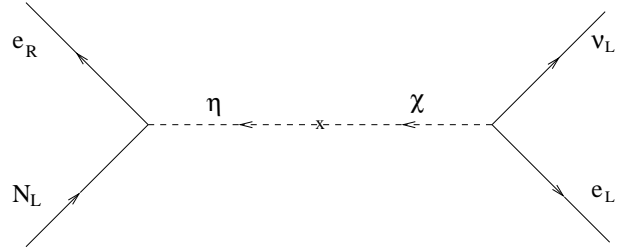, width=0.45\textwidth}
\caption{Scattering of $N_L$ and $e_R$ through the mixing between $\eta^-$ and $\chi^-$.}
\label{scattering_fig}
\end{center}
\end{figure}

Below the mass scale of lightest $\eta^-$, the interaction $h_{ji}^* \eta^- \overline{e_{jR}} 
N_{iL}$ is already gone out-of-thermal equilibrium. Therefore, there is no direct transfer of 
B-L asymmetry stored in $N_L$ to $e_R$. However, in a thermal bath the scattering: 
$\overline{N_{iL}} e_{jR} \rightarrow \nu_{iL} e_{jL}$ can occur through the mixing between 
$\eta^-$ and $\chi^-$ as shown in fig. (\ref{scattering_fig}). Note that this process
violate lepton number by two units and therefore it is suppressed by the
large mass scale of $\eta^-$ and $\chi^-$.

To ensure that, we compute the scattering cross-section at a temperature below the mass 
scale of lightest $\eta^-$:
\begin{equation}
\langle \sigma|v| \rangle =\frac{1}{2 \pi}\frac{|h_{ij}|^2 |f_{ij}|^2 m^4 E^2} 
{M_\eta^4 M_\chi^4}\,,
\end{equation}
where we have assumed that $E_e=E_N=E$. Thus from the above equation we see that below the
mass scale of lightest $\eta^-$ and $\chi^-$, the cross-section goes as: $\langle 
\sigma|v|\rangle \propto m^4/M^8$, assuming that $M_\eta\approx M_\chi=M$. The 
corresponding scattering rate for $N_{iL}$, $i=e, \mu, \tau$ can be given as:
\begin{equation}
\Gamma_{N_L}=n_{N_L}^{eq} \langle \sigma|v| \rangle
\end{equation}
where $n_{N_L}^{\rm eq}$ is the equilibrium number density of $N_L$ and it is given by 
\begin{equation}
n_{N_L}^{eq}=\frac{2 T^3}{\pi^2}
\end{equation}
The out-of-equilibrium of the scattering process: $\overline{N_{iL}}e_{jR} \rightarrow
\nu_{iL}e_{jL}$ then requires
\begin{equation}
\Gamma_{N_L} \lsim H (T\simeq M)\,,
\label{out-of-equilibrium}
\end{equation}
where $H$ is the Hubble expansion parameter. From equation (\ref{out-of-equilibrium})
we get a constraint on the soft B-L violation scale to be 
\begin{equation}
m \lsim 8.54\times 10^9 {\rm GeV}\left( \frac{M}{10^{10} {\rm GeV}}\right)^5 \left(
\frac{\mathcal{O} (1) }{|f_{ij}|^2} \right) \left( \frac{10^{-6}}{ |h_{ij}|^2} \right)\,,
\end{equation}
where we have used $g_*=106$. Thus from the above equation we get a constraint:
$m < {\cal O}(10^{10})$ GeV in order that the scattering will remain in out-of-thermal
equilibrium above the electroweak phase transition.

\subsection{Low reheat temperature and Viability of leptogenesis}
It is believed that the universe has gone through a period of inflation and then reheated to 
a uniform temperature $T_{\rm reh}$. If the corresponding theory of matter is supersymmetric 
then $T_{\rm reh}$ is highly constrained by the success of Big-Bang nucleosynthesis, which 
could be spoiled by the overproduction of gravitino during the radiation dominated 
epoch~\cite{reheating}. 
For a gravitino mass of ${\cal O}$(100 GeV --1 TeV), a conservative upper bound on $T_{\rm reh}$ 
reads $10^{6 -9}$ GeV~\cite{kaz&co}. If $T_{\rm reh}$ is the maximum temperature during 
reheating then it is difficult to create sufficiently high number densities of GUT gauge 
and Higgs bosons including the super heavy $\eta^\pm$ and $\chi^\pm$. However, as discussed in 
refs.~\cite{kolb&turner,riotto&kolb}, after the inflationary era the temperature does not 
rise instantaneously to $T_{\rm reh}$, but rises initially to a maximum temperature 
$T_{\rm max}$ and then falls to $T_{\rm reh}$. It is argued that $T_{\rm max}$ can be 
as high as $10^3 T_{\rm reh}$~\cite{riotto&kolb}. As a result the super-heavy charged particles 
$\eta^\pm$ and $\chi^\pm$ can be easily produced through the gauge interactions: $B_\mu B^\mu 
\rightarrow \eta^+ \eta^-$ and $B_\mu B^\mu \rightarrow \chi^+ \chi^-$. Even though 
$T_{\rm max}$ is quite higher than $T_{\rm reh}$, the gravitinos are mostly produced at 
$T_{\rm reh}$~\cite{raghavan&sahu} and therefore, $T_{\rm max}\gg T_{\rm reh}$ is no more 
harmful for the success of Big-Bang nucleosynthesis. Subsequently the CP violating decay of 
these charged particles can produce lepton asymmetry consistently as discussed above.   

\section{Decaying Dark Matter}
As $N_{1}$ is neutral, it can be a candidate of dark matter. Since $B-L$ symmetry is 
already broken, it can not be a stable dark matter. It will decay through the three 
body process: $N_1 \rightarrow e^- e^+ \nu$, where the neutrino can be in any generation. 
Since this process violate B-L by two unit, it is naturally suppressed by the large mass 
scale of $\chi$ and $\eta$. The decay rate can be given as: 
\begin{equation}
\Gamma_{N_1}=|h_{1e}|^2 |f_{e\tau}|^2\left( \frac{m^2}{M_\eta^2 M_\chi^2} \right)^2 \frac{
M_{N_1}^5}{192 \pi^3}\,, 
\end{equation}
where $h_{1e}$ is the coupling of $N_1$ to $e^+$ and $\eta^-$, while $f_{e\tau}$ is the coupling 
of $\chi^-$ to $e^-$ and $\nu_\tau$. Taking the B-L violating scale $m=10^9$ GeV, 
$M_\eta=3\times 10^{10}$ GeV and $M_\chi=5\times 10^{10}$ GeV, as taken previously, the life time 
of $N_1$ is found to be
\begin{equation}
\tau_{N_1}=0.88 \times 10^{20} {\rm Sec} \left( \frac{{\cal O}(1)}{|f_{e\tau}|^2}\right) 
\left(\frac{10^{-3}}{|h_{1e}|}\right)^2 \left(\frac{10 {\rm MeV}}{M_{N_1}}\right)^5\,.
\end{equation} 
Thus we see that $\tau_{N_1}\gg \tau_0$, where $\tau_0=0.18\times 10^{17}$ Sec, is the age of the 
universe. Thus $N_1$ can be a candidate of dark matter. 

\section{Conclusions}
We proposed an extension of the Zee model with a conserved B-L global symmetry. The
B-L symmetry is softly broken to give rise to the neutrino masses through one loop radiative
correction. In contrast to the Zee model, in the present case the charged scalars
flowing in the neutrino loop are super heavy. Therefore, these charged scalars could
depart from thermal equilibrium in the early universe. As a result the CP violating
decay of the super heavy charged particles, namely $\eta^-$, could generate a net baryon
asymmetry through the leptogenesis route. Recall that the lepton asymmetry is generated
without any B-L violation. This model then accommodate a sterile neutrino like dark matter 
$N_1$ as its three body decay $N_1 \rightarrow e^- e^+ \nu $ is suppressed by the large 
mass scale of $\eta$ and $\chi$.

\section*{Acknowledgement}
NS was supported by the European Union through the Marie Curie Research and Training 
Network "UniverseNet" (MRTN-CT-2006-035863) and by STFC (PPARC) Grant PP/D000394/1. He 
wishes to thank John McDonald for useful discussion.


\begin{thebibliography}{99b}

\bibitem{sahu&sarkar} 
N.~Sahu and U.~Sarkar,
  Phys.\ Rev.\  D {\bf 76}, 045014 (2007);
J.~McDonald, N.~Sahu and U.~Sarkar,
  JCAP {\bf 0804}, 037 (2008)
  [arXiv:0711.4820 [hep-ph]]\,.

\bibitem{neutrino_dark_lep} An incomplete list:
T.~Hambye, K.~Kannike, E.~Ma and M.~Raidal,
  Phys.\ Rev.\  D {\bf 75} (2007) 095003
  [arXiv:hep-ph/0609228];
P.~H.~Gu and U.~Sarkar,
  arXiv:0712.2933 [hep-ph];
E.~Ma, Phys.\ Rev.\ Lett. {\bf 81}, 1171(1998); Phys. Rev.\ D\ {\bf 73}, 077301 (2006);
Mod. Phys. Lett.\ {\bf A21}, 1777 (2006);
N.~Sahu,
  AIP Conf.\ Proc.\  {\bf 939}, 294 (2007)
  [arXiv:0706.0948 [hep-ph]];
N.~Sahu and U.~A.~Yajnik,
  Phys.\ Lett.\  B {\bf 635}, 11 (2006)
  [arXiv:hep-ph/0509285];
K.~S.~Babu and E.~Ma,
  arXiv:0708.3790 [hep-ph].


\bibitem{solar_data} Q.R.~Ahmed {\it et al} (SNO Collaboration), Phys.\ Rev.\ Lett.
{\bf 89}, 011301-011302 (2002);
J.N.~Bahcall and C.~Pena-Garay, [arXiv:hep-ph/0404061].

\bibitem{atmos_data} S.~Fukuda {\it et al} (Super-Kamiokande
Collaboration), Phys.\ Rev.\ Lett. {\bf 86}, 5656 (2001).

\bibitem{kamland_data} K.~Eguchi {\it et al} (KamLAND collaboration),
Phys.~Rev.~Lett. {\bf 90}, 021802 (2003).

\bibitem{rotation_curve} A.~Borriello and P.~Salucci, Mon. Not. Roy. Astron. Soc.
{\bf 323}, 285 (2001)\,.

\bibitem{lensing} J.A.~Tyson, G.P.~Kochanski and I.P.~Dell'Antonio, Astrophys, J. {\bf 498},
L107 (1998); H.~Dahle, [arXiv:astro-ph/0701598]; D.~Clowe, {\it et. al.},
[arXiv: astro-ph/0608407]\,.

\bibitem{structure} M.~Tegmark {\it et. al.}[SDSS Collaboration], Astrophys. J. {\bf 606},
702 (2004).

\bibitem{wmap} D.N.~Spergel {\it et. al.}  D.N.~Spergel {\it et al}
Astrophys.J.Suppl. 148 (2003) 175 [astro-ph/0302209]\,.

\bibitem{typeI_group} P. Minkowski, Phys. Lett. {\bf B 67}, 421 (1977);
M.~Gell-Mann, P.~Ramond and R.~Slansky
in {\it Supergravity} (P.~van Niewenhuizen and D.~Freedman, eds),
(Amsterdam), North Holland, 1979; T.~Yanagida in {\it Workshop
on Unified Theory and Baryon number in the Universe} (O. Sawada
and A.~Sugamoto, eds), (Japan), KEK 1979; R.N.~Mohapatra and
G.~Senjanovic, Phys.\ Rev.\ Lett. {\bf 44}, 912 (1980).

\bibitem{typeII_group} T.~P.~Cheng and L.~F.~Li,
  Phys.\ Rev.\  D {\bf 22}, 2860 (1980);
J. Schechter and J.W.F. Valle, Phys. Rev. {\bf D 22},
2227 (1980); M.~Magg and C.~Wetterich, Phys.~Lett.~B {\bf 94},~61 (1980);
R.~N.~Mohapatra and G.~Senjanovic, Phys. Rev. D{\bf 23}, 165 (1981);
G.~Lazarides, Q.~Shafi and C.~Wetterich, Nucl. Phys.~B{\bf 181}, 287 (1981)\,.

\bibitem{radiative_models} A.~Zee,
  Phys.\ Lett.\  B {\bf 93}, 389 (1980)
  [Erratum-ibid.\  B {\bf 95}, 461 (1980)];
K.~S.~Babu,
  Phys.\ Lett.\  B {\bf 203}, 132 (1988);


\bibitem{maraidalsarkar.99} E. Ma, M. Raidal and U. Sarkar, Phys. Lett. B 460 (1999) 359\,.

\bibitem{lindner} K. Dick, M. Lindner, M. Ratz and D.Wright, Phys. Rev.
Lett. {\bf 84}, 4039 (2000).

\bibitem{ghs7} P.H. Gu and H.J. He, JCAP \textbf{0612}, 010 (2006);
P.H. Gu, H.J. He, and U. Sarkar, JCAP \textbf{0711},
016 (2007); Phys. Lett. \textbf{B 659}, 634 (2008).

\bibitem{petcov}  S.~T.~Petcov,
  Phys.\ Lett.\  B {\bf 115}, 401 (1982).


\bibitem{ma&sarkar} E.~Ma and U.~Sarkar,
  Phys.\ Rev.\ Lett.\  {\bf 80}, 5716 (1998)
  [arXiv:hep-ph/9802445].

\bibitem{reheating}  M. Yu. Khlopov and A. D. Linde,
Phys. Lett. B  {\bf 138} 265 (1984); J. Ellis, J. E. Kim and D. V. Nanopoulos,
Phys. Lett. B {\bf 145} 181 (1984); R. Juszkiewicz, J. Silk and A. Stebbins,
Phys. Lett. B{\bf 158} 463 (1985); J. Ellis, D. V. Nanopoulos and S. Sarkar,
Nucl. Phys. B {\bf 259} 175 (1985); M. Kawasaki and K. Sato,
Phys.\ Lett.\ B{\bf 189} 23 (1987); M. Yu. Khlopov, Yu. L. Levitan, E. V. Sedel'nikov, and
I. M. Sobol, 
Phys. Atom. Nucl.
{\bf 57} 1393 (1994) [Yad.\ Fiz.\  {\bf 57}, 1466 (1994)]; 
 M.~Kawasaki and T.~Moroi, 
Prog.\ Theor.\ Phys.\  {\bf 93}, 879 (1995)
[hep-ph/9403364]; 
R. H. Cyburt, J. R. Ellis, B. D. Fields and K. A. Olive,
Phys. Rev. D {\bf 67} 103521 (2003)\,.

\bibitem{kaz&co} M.~Kawasaki, K.~Khori and T.~Moroi, Phys.\ Rev.\ D{\bf 71}, 083502 (2005)\,.

\bibitem{kolb&turner} E.W.~Kolb and M.S.~Turner, {\it The Early Universe}, Addison-Wesley, 1990\,.

\bibitem{riotto&kolb} D.J.H.~Chung, E.W.~Kolb and A.~Riotto, Phys.\ Rev.\ D {\bf 60}, 063504 (1999)\,.

\bibitem{raghavan&sahu}
R.~Rangarajan and N.~Sahu,
  Mod.\ Phys.\ Lett.\  A {\bf 23}, 427 (2008)
  [arXiv:hep-ph/0606228].

\end{thebibliography}
\end{document}